\documentclass{aa}  

\usepackage{subfig}
\usepackage{graphicx}
\usepackage{pifont}
\usepackage{txfonts}
\newcommand{\degree}{^{\circ}}
\begin{document}

   \title{Injection of solar energetic particles into both loop legs of a magnetic cloud}

\titlerunning{The 7~Nov~2013 SEP event}

   \author{N. Dresing
          \inst{1}
          \and
          R. G\'omez-Herrero\inst{2} 
          \and
          B. Heber\inst{1}
          \and
          M.A. Hidalgo\inst{2}
          \and
          A. Klassen\inst{1}         
          \and
          M. Temmer\inst{3} 
                  \and
                  A. Veronig\inst{3}
          }

   \institute{Institut f\"ur Experimentelle und Angewandte Physik, University of Kiel, Germany\\
   \email{dresing@physik.uni-kiel.de}   
         \and
             Space Research Group, Dpto. de F\'isica y Matem\'aticas, University of Alcal\'a, Spain\\   
          \and
             Institute of Physics/Kanzelhöhe Observatory, University of Graz, Austria\\  
             }
             
   \date{}

 
  \abstract
{Each of the two Solar TErrestrial RElations Observatory (STEREO) spacecraft carries a Solar Electron and Proton Telescope (SEPT) which measures electrons and protons. Anisotropy observations are provided in four viewing directions: along the nominal magnetic field Parker spiral in the ecliptic towards the Sun (SUN) and away from the Sun (Anti-Sun / ASUN), and towards the north (NORTH) and south (SOUTH).
The solar energetic particle
(SEP) event on 7 November 2013 was observed by both STEREO spacecraft, which were longitudinally separated by 68$\degree$ at that time. 
While STEREO~A observed the expected characteristics of an SEP event at a well-connected position, STEREO~B detected a very anisotropic bi-directional distribution of near-relativistic electrons and was situated inside a magnetic-cloud-like structure during the early phase of the event.
}
%
{
We  examine the source of the bi-directional SEP distribution at STEREO~B.
On the one hand this distribution could be caused by a double injection into both loop legs of the magnetic cloud (MC).
On the other hand, a mirroring scenario where the incident beam is reflected in the opposite loop leg could be the reason.
Furthermore, the energetic electron observations are used to probe the magnetic structure inside the magnetic cloud.
}
%
%
{We investigate in situ plasma and magnetic field observations and show that STEREO~B was embedded in an MC-like structure ejected three days earlier on 4 November from the same active region.
We apply a Graduated Cylindrical Shell (GCS) model to the coronagraph observations from three viewpoints as well as the Global Magnetic Cloud (GMC) model to the in situ measurements at STEREO~B to determine the orientation and topology of the MC close to the Sun and at 1\,AU. 
We also estimate the path lengths of the electrons propagating through the MC to estimate the amount of magnetic field line winding inside the structure.
}
{The relative intensity and timing of the energetic electron increases in the different SEPT telescopes at STEREO~B strongly suggest that the bi-directional electron distribution is formed by SEP injections in both loop legs of the MC separately instead of by mirroring farther away beyond the STEREO orbit.
Observations by the Nancay Radioheliograph (NRH) of two distinct radio sources during the SEP injection further support the above scenario.
The determined electron path lengths are around 50\% longer than the estimated lengths of the loop legs of the MC itself (based on the GCS model) suggesting that the amount of field line winding is moderate.}
{}


   \maketitle
%

\section{Introduction}
The acceleration, injection, and propagation of solar energetic particles (SEPs) from their source to an observer in interplanetary space have been investigated over the last decades.
While multi-spacecraft observations go back to the 1970s, modern capabilities of remote sensing observations from several viewpoints around the Sun in combination with many in situ measurements provide essential input to establish a comprehensive picture of the physics of SEP events.
The Solar TErrestrial RElations Observatory (STEREO) mission launched in October 2006 provides an unprecedented platform for investigating SEP events from multiple points in space.
While radial gradient effects can be neglected owing to the similar heliocentric distance of the spacecraft and Earth, the increasing angular separation between the two spacecraft offers the the ability to investigate the longitudinal intensity distribution of such events and the discovery of so-called wide-spread SEP events \citep{Dresing2012, Lario2013, Dresing2014, Richardson2014, Gomez-Herrero2015}.
In February 2011 the spacecraft reached a separation of $180\degree$ from each other and in March 2015 they were in superior conjunction on the opposite side of the Sun as seen from Earth.
In this constellation detailed multipoint analyses of events are now possible especially on small spatial scales \citep{Klassen2015} . \\
In the present study we report on the SEP event of 7 November 2013 where energetic electrons and protons were observed by both STEREO~A (STA) and STEREO~B (STB), which were separated by $68\degree$ in heliographic longitude.
While STA detects a beamed distribution with SEPs streaming from the Sun, which is usual for a well-connected observer, STB observes a bi-directional distribution in the north-south direction.
Bi-directional particle flows may be caused by different processes, which can be split into two  scenarios: i) Particle injections from two sides (which may be different particle sources) or ii) an incident particle beam that is mirrored.
The mirroring may take place behind the observer in an open magnetic field regime.
In this case the magnetic mirror can be created by various structures in the heliosphere such as a corotating interaction region (CIR), a magnetic cloud or interplanetary coronal mass ejection (ICME), a shock, or turbulence \citep{Bieber2002, Roelof2008}.
On the other hand, mirroring can also happen inside a closed magnetic structure such as a magnetic cloud building a magnetic bottle configuration at each loop leg,  still anchored at the Sun.
In this special situation an observer situated inside a magnetic cloud may observe an incident beam injected into one loop leg of the structure followed by a secondary beam that was mirrored in the opposite loop leg.
The solar wind suprathermal electron heat flux then shows bi-directional flows and can be taken as a marker for the loop legs being rooted at the Sun.\\
Sometimes SEPs are injected into large-scale closed magnetic structures such as magnetic clouds (MCs).
Because these structures may significantly change the interplanetary magnetic field topology, a good connection to the east or to the far west   (behind the west limb) may appear.
Impulsive SEP events from these regions can then be observed \citep{Richardson1991, Gomez-Herrero2006}.
Without such a structure the observer would be nominally connected to $\sim$~W60 (assuming a solar wind speed of 400\,km/s) and the SEP event from a far separated source region would be expected to be more gradual owing to perpendicular transport or would even not be observed.\\
%
Several SEP events inside magnetic flux ropes showing bi-directional SEP distributions have been analyzed \citep[e.g.,][]{Richardson1996, Torsti2004, Leske2012}.
However, usually the cause of the bi-directional streaming cannot be identified unambiguously.
Making a clear distinction between a mirroring scenario or an injection into both loop legs is usually very complicated.
It is likely that often a mixture of both a double injection and mirroring is present.
Proof of a mirrored beam could be the presence of a loss cone distribution, which can only be observed if a high pitch angle resolution of the beam is available.
To our knowledge a double-injection scenario could never be proven unambiguously, although it was often presumed \citep[e.g.,][]{Richardson1996, Leske2012}.
Only modeling results of neutron monitor measurements of the 22 October 1989 event by \cite{Ruffolo2006} strongly suggested such a scenario.
\\
In this paper we will show that STB was embedded in a north-south oriented MC-like structure during the early phase of the 7 November 2013 SEP event, which was most likely not the case for STA.
In Section \ref{sec:obs} we analyze remote sensing as well as in situ plasma, magnetic field, and energetic particle observations of the 7 November 2013 SEP event to determine whether the bi-directional SEP distribution is formed by  a mirroring effect or by an injection into both loop legs.
The results of these observations are discussed in Section \ref{sec:disc_inj}.
The onset timing and relative intensities of the two beams reaching STB strongly suggest that the bi-directional electron distribution is produced by an injection into both loop legs of the magnetic cloud structure rather than a mirroring effect.
Together with the STA measurements of the same event we put the observations into a multipoint context to gain information on the SEP injection close to the Sun and the transport through the IP magnetic field.\\
In Section \ref{sec:mag_cloud} we analyze in detail the MC which was launched on 4 November and embeds STB during the 7 November event.
After discussing the observations in Section \ref{mc_obs} we apply a Graduated Cylindrical Shell (GCS) Model \citep{Thernisien2006, Thernisien2009} to the coronagraph observations and fit the in situ flux rope observations \citep{Hidalgo2014} to determine the topology and orientation of the MC in the interplanetary (IP) medium.
In Section \ref{sec:results_mc} the results of the MC observations and modeling are discussed.
Finally we combine the above findings in Section \ref{disc2:twist} and use the calculated electron path lengths (determined in Section \ref{sec:path_length}) in both MC loop legs to probe the structure of the MC.
We find that the winding of the magnetic field lines at the borders of the MC, which is predicted by models \citep[e.g.,][]{Lepping1990}, must be moderate in the analyzed case.
%
\section{Instrumentation}\label{sec:instr}
Both STEREO spacecraft carry  remote sensing technology and in situ instruments.
The EUV telescope (EUVI, \cite{Wuelser2004}) and the coronagraphs (COR1, COR2) are contained in the SECCHI investigation \citep{Howard2008}.
The magnetometer \citep{Acuna2007} and the energetic particle experiments such as the High Energy Telescope (HET, \cite{Rosenvinge2008}, the Low Energy Telescope (LET, \cite{Mewaldt2007}), and the Solar Electron and Proton Telescope (SEPT, \cite{Muller-Mellin2007}) are contained in the IMPACT instrument suite \citep{Luhmann2007}.
Only two energetic particle detectors provide directional measurements: LET, which measures protons and heavier ions, and SEPT, which  measures electrons in the energy range of 45-400\,keV and ions in the range of 60-7000\,keV.
Each SEPT consists of two units covering both forward and backward directions with a telescope opening angle of $52.8\degree$ for electrons.
Each STEREO spacecraft carries two SEPT instruments  that are mounted perpendicularly to each other to provide four viewing directions.
One SEPT is mounted in the ecliptic plane tilted by $45\degree$ against the Sun-spacecraft line so that it points along the nominal magnetic field spiral.
The telescope oriented along the field line towards the Sun is called SUN; the other one, which points away from the Sun, is called ANTI-SUN.
SEPT also covers the northern and southern hemispheres, in contrast to LET, which  only provides sectored measurements centered in the ecliptic plane spanning a field of view of 133$\degree$ in longitude and $29\degree$ in latitude. 
Therefore, the second SEPT unit is mounted perpendicularly to the first one and looks north (NORTH) and south (SOUTH).
Sectored intensity measurements are indispensable to characterize the energetic particle distribution in terms of its incident direction, the anisotropy (i.e., how beamed the distribution is) and the shape of the distribution (e.g., a beam, a bi-directional, or an isotropic distribution). \\
Solar wind plasma measurements are provided by the PLASTIC instrument \citep{Galvin2008}.
Signatures of interplanetary radio bursts are detected with the STEREO/SWAVES instruments \citep{Bougeret2008}.
We complement the STEREO observations with ground-based radio observations from the Nancay Radioheliograph (NRH, \cite{Kerdraon1997}) and measurements by the WAVES instrument aboard Wind \citep{Bougeret1995}, as well as coronagraph observations by LASCO \citep{Brueckner1995} aboard SOHO.
%
\section{Observations of the 7 November SEP event}\label{sec:obs}
%
\subsection{Remote sensing observations}\label{sec:remote_obs}
\begin{figure}[h!] 
\centering{
\includegraphics[width=0.45\textwidth, clip=true, trim = 10mm 10mm 10mm 0mm]{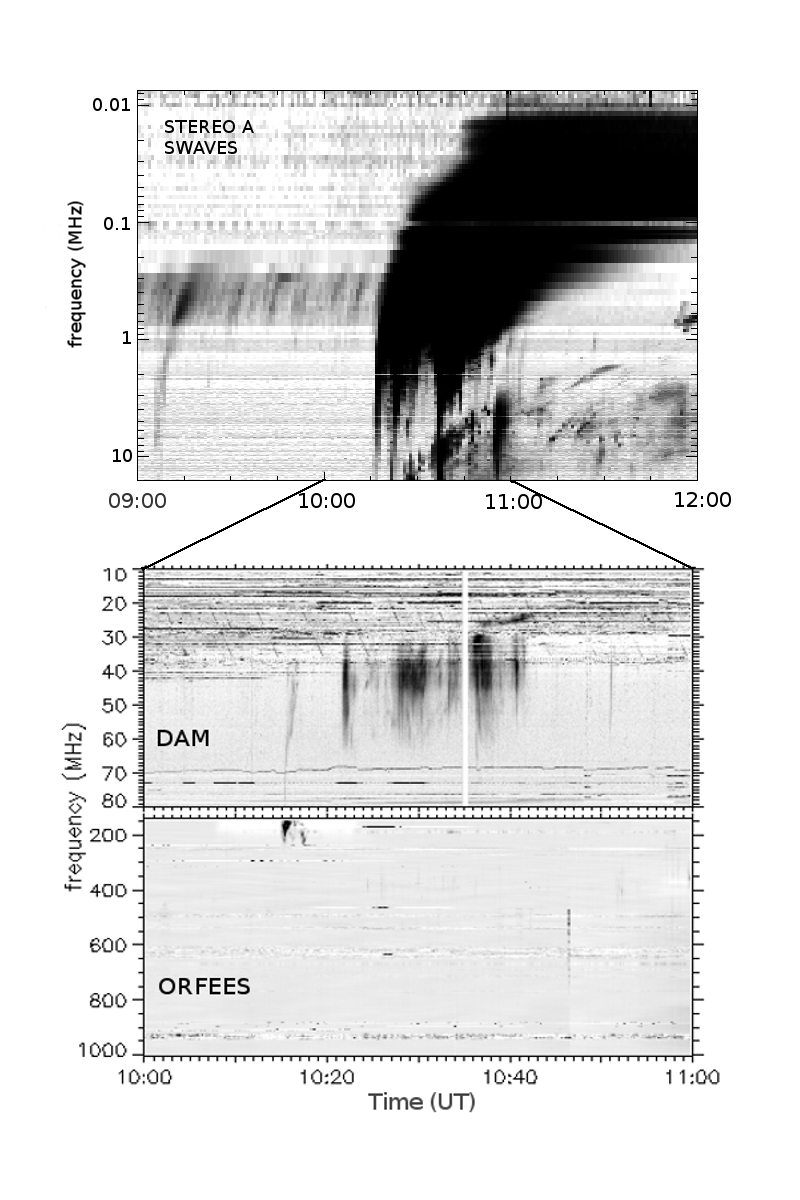}}
 \caption{Radio spectrograms recorded by the SWAVES instruments on board STA (top) from 09:00 until 12:00\,UT. The bottom two panels  show radio spectra detected by the ground-based stations DAM (middle) and ORFEES (bottom) from 10:00 to 11:00\,UT. Adopted from the Meudon Radio Monitoring Survey (http://secchirh.obspm.fr/select.php).} \label{fig:radio}
\end{figure}%
\begin{figure}[h!] 
\centering{
\includegraphics[width=0.35\textwidth, clip=true, trim = 0mm 0mm 0mm 0mm]{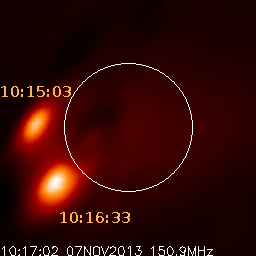}}
 \caption{Two distinct radio sources as observed by NRH at 150.9\,MHz provided by the Meudon Radio Monitoring Survey (http://secchirh.obspm.fr/select.php). The time of the first appearance of each of the two radio sources is given in the figure.} \label{fig:radio2sources}
\end{figure}%
On 7 November 2013 active region (AR) 11875 produced a flare starting around 10:10\,UT to 10:15\,UT (observed by both STEREO/EUVI with five minute cadence) at the backside of the Sun as seen from Earth.
The flare occurred at $+3\degree$ latitude and $37\degree$ Carrington longitude (E152 with respect to the Earth) situated between the nominal magnetic footpoints of both STEREO spacecraft which were separated by $68 \degree$ in longitude at that time (see Fig. \ref{fig:sketch}).
The longitudinal separation between STA (STB) and the Earth was 143$\degree$ (149$\degree$).
A series of type III radio bursts starting at 10:15\,UT and a type II burst marking the presence of a coronal shock were observed by all three vantage points (see Fig. \ref{fig:radio} showing STA/SWAVES and ground-based  observations).
The first signatures of the type II burst are observed by STEREO at 10:25\,UT at 16\,MHz.\\
Although the flaring AR of the 7 November event is $62\degree$ behind the east limb as seen from Earth, the ground-based radio observatory NRH observes in temporal coincidence with the flare two distinct radio sources above the east limb at 150.9\,MHz (Fig. \ref{fig:radio2sources}).
These sources simultaneously last for at least 2 minutes. 
As indicated in Fig. \ref{fig:radio2sources}, the northern source appears first at 10:15:03 UT, the southern one 90 seconds later at 10:16:33 UT.
Each of these two radio sources was associated with a group of type III-like radio bursts limited in frequency and detected around 150\,MHz (see Fig. \ref{fig:radio}, ORFEES).
At least one of them was also observed at frequencies below the NRH range (at 80-40\,MHz, see Fig. \ref{fig:radio}, DAM) and  below 16\,MHz (Fig. \ref{fig:radio}, SWAVES) when the first type III burst at 10:15\,UT appears and is detected at all three spacecraft (not shown).
This suggests that both radio sources mark energetic electron beams injected into two different magnetic field structures.
\\
Later from 10:22\,UT until 10:42\,UT a first signature of type II radio emission (mainly herringbones) appears in the frequency range 60-30\,MHz with prolongation into the SWAVES range below 16\,MHz.
At the same time some shock accelerated/associated type III bursts obviously escape from the type II lane (e.g., 10:22\,UT) and are seen in the SWAVES frequency range (Fig. \ref{fig:radio} top).
This implies a possible second source producing energetic electrons injected into the IP medium due to the shock.\\
The flare on 7 November is accompanied by an EIT wave and a CME which is listed in the CACTUS and LASCO CME catalogs (\url{http://sidc.oma.be/cactus/}, \url{http://cdaw.gsfc.nasa.gov/CME_list/}).
STA has the most advantageous perspective onto the CME from the side and a CME width of $276\degree$ is listed in the CACTUS STA catalog.
The CME speed as estimated from the different viewing perspectives ranges between 1041\,km/s (CACTUS STA) and 1405\,km/s (LASCO, linear speed).
The associated ICME is listed in the STEREO ICME catalog compiled by Lan Jian (UCLA/IGPP) with an arrival time of the magnetic obstacle  at STB at 18:44\,UT on 8 November.
The criteria for event selection are given in \cite{Jian2006a, Jian2013}.
It is not observed in situ  at STA.
%
\subsection{In situ Observations}\label{sec:in_situ}
\begin{figure}[h!] 
\centering{
\includegraphics[width=0.4\textwidth, clip=true, trim = 20mm 10mm 60mm 130mm]{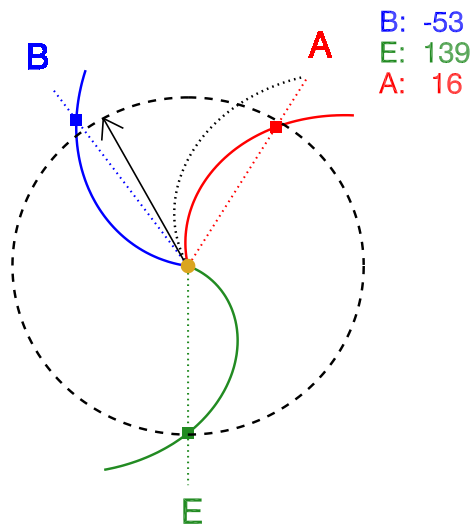}}
 \caption{Longitudinal configuration of STA (red), STB (blue), and the Earth (green). The spirals denote the nominal Parker field lines corresponding to the measured solar wind speed. The black arrow indicates the longitude of the flare and the dotted spiral represents the nominal connecting field line to the flare longitude assuming a solar wind speed of 400\,km/s. The longitudinal separation angles between the backmapped magnetic footpoint of the spacecraft and the flare are given in the top right corner of the figure. Negative numbers denote a footpoint lying to the west of the flare's longitude.} \label{fig:sketch}
\end{figure}

\begin{figure*}[t] 
\begin{minipage}{.49\linewidth}
\centering
\subfloat[]{\includegraphics[width=\textwidth, clip=true, trim = 0mm 15mm 0mm 0mm]{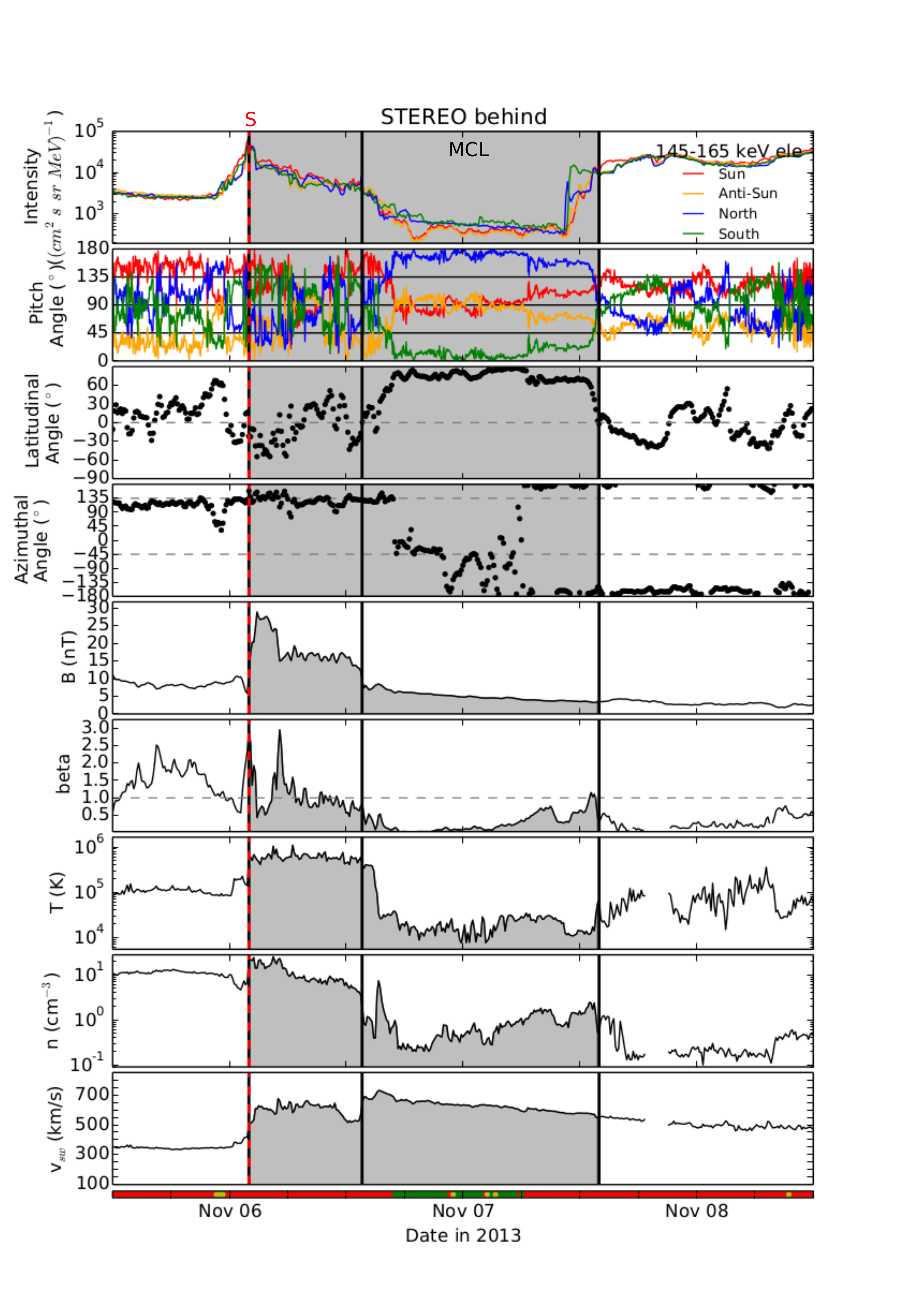}}
\end{minipage}%
\begin{minipage}{.49\linewidth}
\centering
\subfloat[]{\includegraphics[width=\textwidth, clip=true, trim = 0mm 15mm 0mm 0mm]{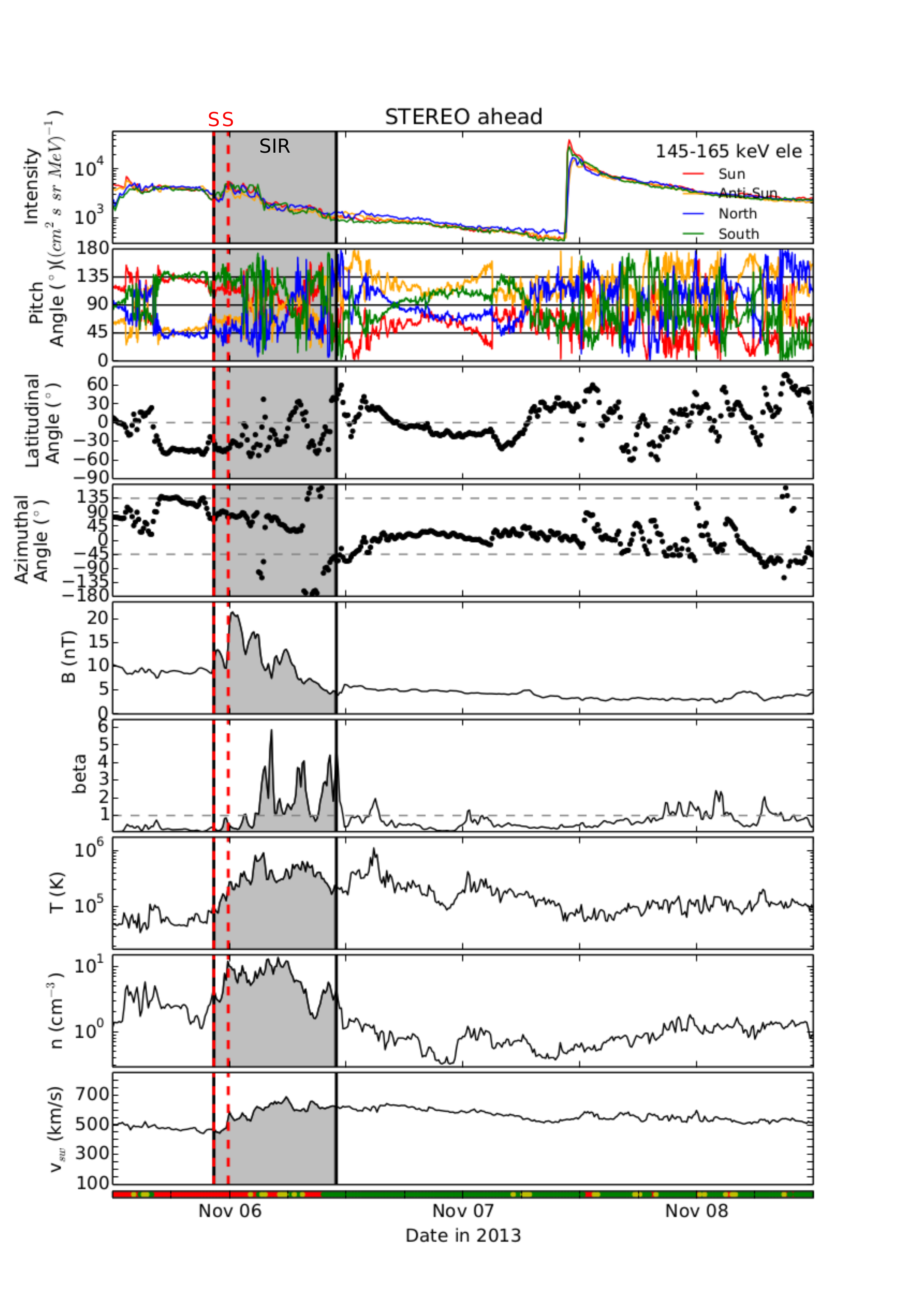}}
 \end{minipage}
 \caption{Solar wind plasma and magnetic field observations from 5 November until 9 November 2013 observed by STB (a) and STA (b). The top panel shows 145-165\,keV electrons observed by SEPT in its four viewing directions. The panel below plots the pitch angles of the four SEPT telescopes. Below the latitudinal and azimuthal angle of the magnetic field in RTN coordinates are shown the magnetic field magnitude, the plasma beta, the solar wind proton temperature, the proton density, and the solar wind speed. The red/green bar at the bottom represents the in situ polarity of the magnetic field with red (green) standing for negative (positive) polarity and yellow meaning uncertain periods. The shaded range in (a) marks the duration of an ICME as listed in the ICME catalog by Lan Jian (\protect\url{http://www-ssc.igpp.ucla.edu/forms/stereo/stereo_level_3.html}) with a magnetic obstacle (identified as a magnetic-cloud-like structure, labeled MCL) between the second and third vertical lines. The red dashed line (labeled  S) indicates the associated CME-driven shock listed in the Jian shock list. The shaded range in (b) indicates a SIR as listed in the Jian SIR catalog. Two shocks (labeled  S, from the Jian catalog) pass STA at the beginning of the SIR.} \label{fig:overview_STB_STA} 
\end{figure*}

The associated solar energetic electrons arrive at both STEREO spacecraft, but no event is observed by  spacecraft close to the Earth, which suggests that this event does not belong to the wide-spread SEP event class.
Fig. \ref{fig:sketch} shows the longitudinal constellation of STA (red), STB (blue), and the Earth (green).
The spirals represent the magnetic field lines connecting the spacecraft with the Sun according to the measured solar wind speeds at the time of the event. 
The black dotted curve represents the magnetic field spiral connected to the flare longitude assuming the same solar wind speed as measured by STA.
The longitude of the flare is indicated by the black arrow.
The longitudinal separation angles between the spacecraft backmapped magnetic footpoints and the flare are shown in the upper right corner of the figure.
With a separation of $16\degree$, STA is nominally well connected, which is not the case for STB with a separation of $53\degree$.
Nevertheless, the associated energetic electron onset is nearly simultaneous at both STEREO spacecraft.\\
\begin{figure}[h!] 
\centering{
\includegraphics[width=0.5\textwidth, clip=true, trim = 0mm 35mm 40mm 0mm]{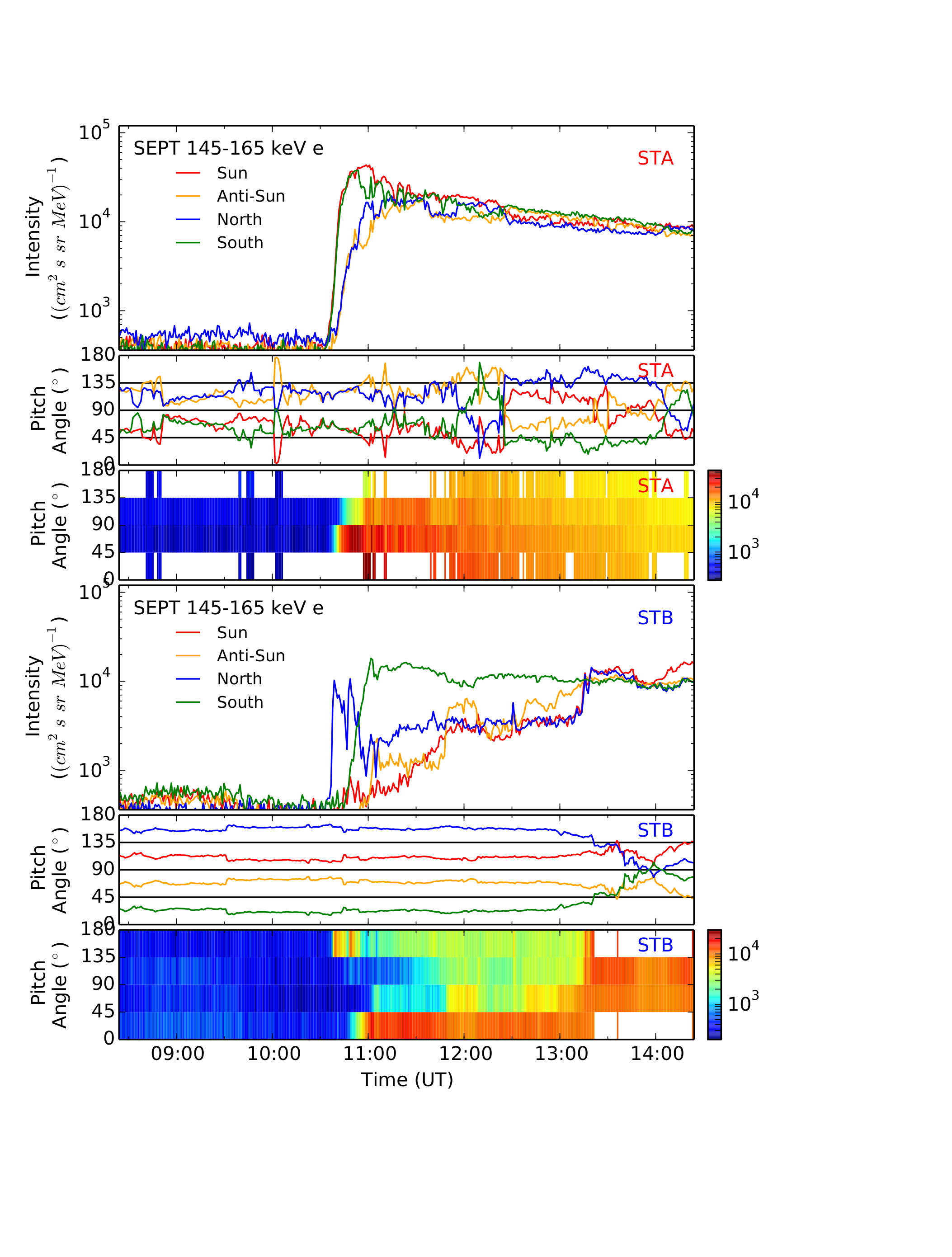}}
 \caption{Energetic electron observations (145-165\,keV) at STA (top three panels) and STB (bottom three panels) on 7 November 2013. In each group of panels, the  uppermost panel presents the electron intensities in the four SEPT telescopes, the middle panel shows the corresponding pitch angles of the four telescopes, and the bottom panel displays the pitch angle distribution with color coding for the intensity.} \label{fig:SEPT_sectored_intro}
\end{figure}%
The uppermost panel of Fig. \ref{fig:SEPT_sectored_intro} displays the intensity of about 155\,keV electrons as measured by the four SEPT telescopes on board STA.
At 10:36\,UT the SUN and SOUTH telescopes observe an intensity increase.
Both telescopes  cover the same pitch angle of $\sim70\degree$ (shown in the second panel of the figure).
The other two SEP telescopes (ASUN and NORTH) cover pitch angle $\sim120\degree$ (meaning 120 degrees away from the guiding magnetic field line) and observe an onset at 10:42\,UT.
Although the pitch angle coverage is not ideal for the first hour of the event, a clear anisotropy can be identified.
The increase shows a beamed distribution of particles streaming away from the Sun as expected for an SEP event.
The onset time at pitch angle 0$\degree$ (along the guiding magnetic field line, which is not covered) is therefore expected to be at least one minute earlier.
The third panel presents the pitch angle distribution with the intensity in color coding.
At the time of maximum intensity,  the pitch angle coverage improves for a period of a few minutes;
a range of less than 35$\degree$ up to more than 155$\degree$ is covered (see second panel from top) so that the intensity maximum is observed in the SUN telescope (top panel). \\
The three lower panels show the same for STB, respectively, which has a much better pitch angle coverage during the event (see second panel from bottom).
The first arriving particles are observed in the NORTH telescope at 10:37\,UT (third panel from bottom).
However, a second beam is detected in the SOUTH sector 12 minutes later at 10:49\,UT. 
In the ASUN and SUN telescopes the intensity begins to rise even later and shows a very gradual profile.
The pitch angle distribution observed by STB is a bi-directional distribution (cf. bottom panel).
Surprisingly, despite the later onset in the SOUTH sector, the peak intensity is higher than the NORTH sector, which is discussed in more detail in Section \ref{sec:disc_inj}.
We note that low energy bins $<120$\,keV in the STB/NORTH telescope were possibly contaminated by high energy (>2\,MeV) electrons and were therefore discarded in this investigation.
The presence of higher energy electrons is confirmed by the HET measurements showing electrons in the highest available energy channel of 2.8-4.0\,MeV at both STEREO spacecraft (not shown). 
We also note, however, that the HET viewing cone at STB does not point along the magnetic field line as long as STB is situated inside the MC.
\\
Both STA and STB also detect energetic protons.
A clear event up to the highest available proton channel of the HET instrument at 60-100\,MeV (not shown) is seen by both observers.
The sectored proton measurements of the LET instruments confirm the electron distributions observed by SEPT:
STA detects an anisotropic event with particles streaming away from the Sun.
At STB the LET 4-6\,MeV protons begin to rise when the spacecraft leaves the MC-like structure, which is expected because the viewing directions of LET only cover the ecliptic plane within $\pm15\degree$.
However, the protons still show a bi-directional distribution that lasts for a few hours.\\
Assuming a solar release time at 10:15\,UT (when the first type III burst is observed) the 155\,keV electrons arrive at both STEREO spacecraft with a delay of around 15 minutes with respect to a scatter-free propagation time along a corresponding Parker field line.
\\
The fact that pitch angles 0$\degree$ and 180$\degree$ at STB are covered by the SOUTH and NORTH telescopes during the onset of the event shows that the magnetic field is oriented in the north-south direction.
Fig. \ref{fig:overview_STB_STA} (a) shows an overview of the in situ plasma and magnetic field data observed by STB. 
The two upper panels show again the energetic electron intensities in the four SEPT sectors and the corresponding pitch angles (as in Fig. \ref{fig:SEPT_sectored_intro}).
The latitudinal and azimuthal angles of the magnetic field are plotted in the two panels below, followed by the magnetic field magnitude, the plasma beta, the solar wind proton temperature, proton density, and the solar wind speed.
The colored bar below the bottom panel represents the in situ magnetic field polarity with red marking negative, and green positive polarity, respectively.
Yellow stands for uncertain periods.
The SEPT pitch angles (in the second panel from top) and the magnetic field angles clearly show that STB is embedded in a magnetic structure with a north-south oriented magnetic field when the SEP event on 7 November occurs.
The magnetic field direction is very stable with only very few fluctuations over nearly a day.
The gray-shaded range marks the ICME duration as listed in the ICME catalog by Jian (\url{http://www-ssc.igpp.ucla.edu/forms/stereo/stereo_level_3.html}).
The magnetic obstacle of this ICME is labeled MCL (MC-like structure), the red vertical line on the left marks the CME-driven shock (labeled S, listed in the shock catalog by Jian, see above).
The magnetic obstacle is accompanied by low proton temperatures and densities, and a smooth magnetic field showing some rotation and a low plasma beta denoting that the structure is magnetically dominated.
However, to be classified as a magnetic cloud \citep{Burlaga1981} an enhanced magnetic field strength is also required, which is not clearly observed.
Although the local observations at STB show only signatures of a {\it MC-like} structure, we will call the structure a MC  in the following when its overall magnetic structure in the IP medium is discussed.
The topology and kinematics of the CME and embedded MC is discussed in more detail in Section \ref{sec:results_mc}.
A clear bi-directional streaming of the electron heat flux is observed by the STB/SWEA instrument (\url{http://stereo.cesr.fr/indexplots.php}, not shown) in close association with the magnetic cloud-like structure seen in the plasma magnetic field observations.
This bi-directional heat flux is usually interpreted as a signature of a closed magnetic structure, i.e., with loop legs anchored at the Sun \citep{Larson1997, Bothmer1998}.\\
Figure \ref{fig:overview_STB_STA} (b) shows the same in situ measurements as observed by STA.
The gray-shaded range marks a stream interaction region (SIR) as listed in the SIR catalog by Jian (see above).
The two dashed red lines mark two forward shocks also provided by the Jian shock catalog.
Although STA is most likely not embedded in the same magnetic structure  present at STB, it should be noted that STA observes a bi-directional electron heat flux  (from 6 November $\sim$4:40\,UT until 7 November $\sim$19:00\,UT, not shown), which is less intense than that observed by STB perhaps because of a reflecting boundary behind the spacecraft, possibly provided by the preceding SIR.
A close inspection of the energetic electrons observed by STA does not show any evidence of a bidirectional streaming or a less intense mirrored electron beam.
Therefore, we believe that STA is situated in an open magnetic field structure during the 7 November SEP event while STB is embedded in a magnetic-cloud like structure with its looplegs still anchored at the Sun.
%
\section{Results I: SEP injection and propagation in the magnetic cloud observed by STEREO~B}\label{sec:disc_inj}
If an SEP event is injected into an ideal closed magnetic flux-rope structure we usually expect a bi-directional SEP distribution inside the structure because of the mirroring effect at the loop legs where the magnetic field converges.
Depending on the effectiveness of the mirror, the intensity of the reflected beam may be higher or lower.
Values of up to 30\% for near-relativistic electrons were observed  by e.g., \cite{Agueda2010} and \cite{Wang2011} (see also \cite{Klassen2012}). 
In either case the incident beam is expected to be more intense and should arrive earlier than the mirrored beam.
In the case of the 7 November 2013 SEP event STB observes a bi-directional electron distribution in the NORTH and SOUTH telescopes as discussed in Section \ref{sec:obs} (see Fig. \ref{fig:SEPT_sectored_intro}).
The first arriving beam is observed in the NORTH telescope suggesting that this beam was directly injected into the MC.
The SOUTH telescope detected an increase 12 minutes later than the onset in the NORTH telescope.
Therefore, this later beam could either be the mirrored part of the incident beam or it could be a distinct injection into the other loop leg.
The delay of 12 minutes is too short for a mirroring scenario {close to the Sun,} which should take about 30 minutes assuming a total path length of 2.3\,AU traveled by 155\,keV electrons.
On the other hand, it is not clear at what distance the mirroring point would be.
A stronger point against the mirroring scenario is that the later arriving beam in the SOUTH telescope is {more intense} than the earlier beam detected in the NORTH telescope.
Because the pitch angle coverage of SEPT is sufficient and stable during the rising phase of the event, the intensity and timing differences are reliable.
Therefore, the observations strongly suggest that the later arriving beam is not formed by mirroring but by a distinct injection, meaning that SEPs must have been injected into both loop legs of the MC-like structure separately.
The different shapes of the intensity time profiles in the NORTH and SOUTH telescopes, as well as the long-lasting anisotropy, are in agreement with the above assumption.
The above scenario is further supported by the two distinct radio sources observed by NRH (see Section \ref{sec:remote_obs} and Fig. \ref{fig:radio2sources}).
The presence of these two separate sources fits well with the scenario of an energetic particle injection into two separated loop legs of the magnetic cloud-like structure.\\
In addition to  the flare, the 7 November event is associated with a coronal and IP shock as shown by the observations of a type II radio burst (see Fig. \ref{fig:radio}).
This shock may act as an extended source region in the corona. 
An extension of the shock towards the west (east limb as seen from Earth) may be indicated by the low frequency part of the type II burst that is observed from the Earth (ground-based) and by Wind, although the associated flaring AR is situated 62$\degree$ behind the limb.\\
Such an extended source region together with the flare source may provide SEP injections into open field lines towards STA as well as into both loop legs of the magnetic cloud comprising STB.
\begin{figure*}[t!] 
\centering{
\includegraphics[width=0.85\textwidth, clip=true, trim = 0mm 0mm 0mm 0mm]{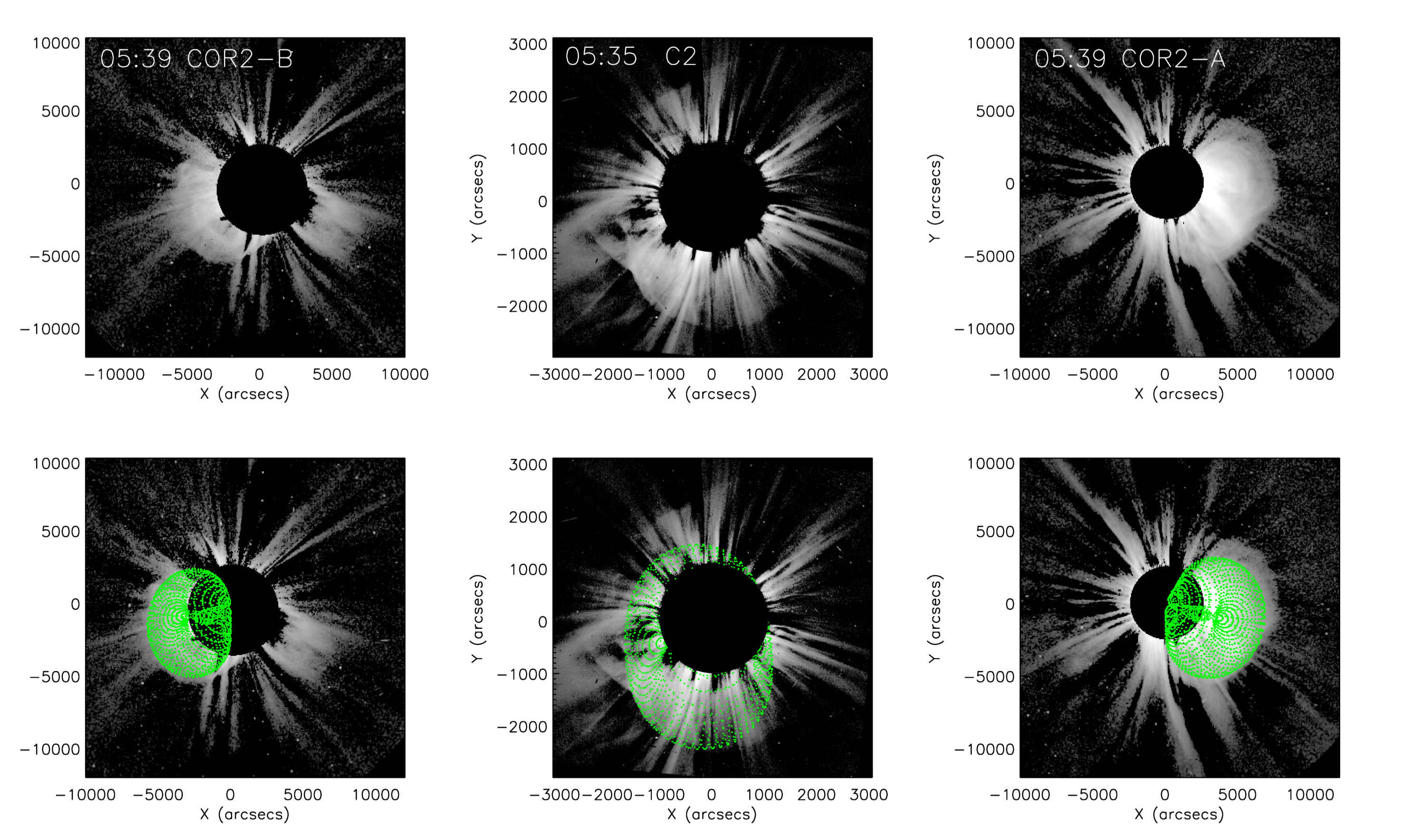}}
 \caption{Graduated Cylindrical Shell (GCS) model to reproduce the orientation and topology of the MC launched on 4 November 2013. The top panel shows coronagraph observations by STB/COR2 (left), SOHO/LASCO/C2 (middle), and STA/COR2 (right). The lower panel shows the same with an overplotted grid showing the GCS model.} \label{fig:gcs_model}
\end{figure*}%
\begin{figure}[hb!] 
\centering{
\includegraphics[width=0.48\textwidth, clip=true, trim = 25mm 20mm 25mm 20mm]{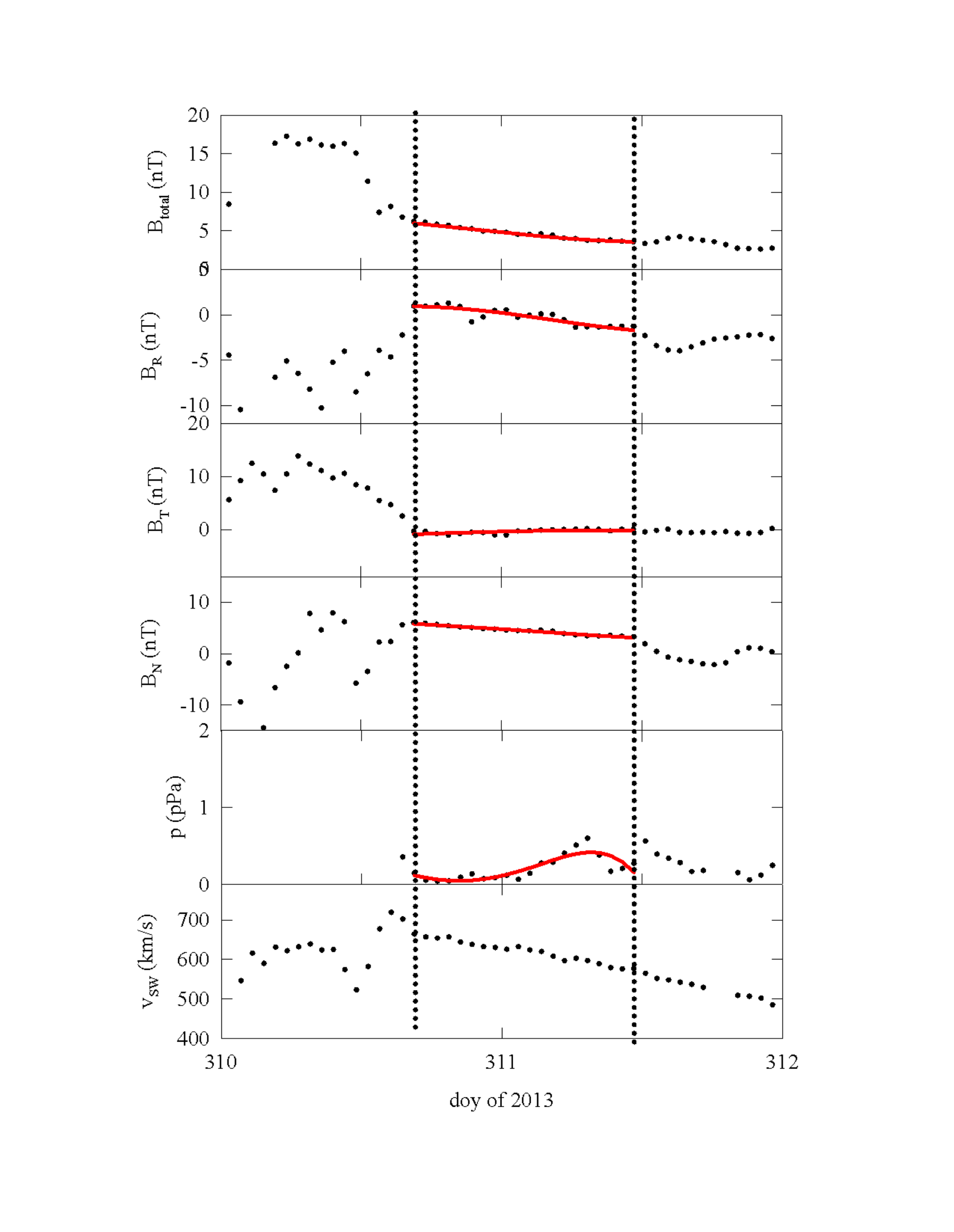}}
 \caption{In situ measurements at STB (dots) and fits of the GMC model (red lines). From top to bottom: Magnetic field magnitude, the three RTN magnetic field components, plasma pressure, and solar wind speed.} \label{fig:mc_model}
\end{figure}%
\begin{figure}[b!] 
\begin{minipage}{\linewidth}
\centering
\subfloat[]{\includegraphics[width=\textwidth, clip=true, trim = 0mm 0mm 0mm 0mm]{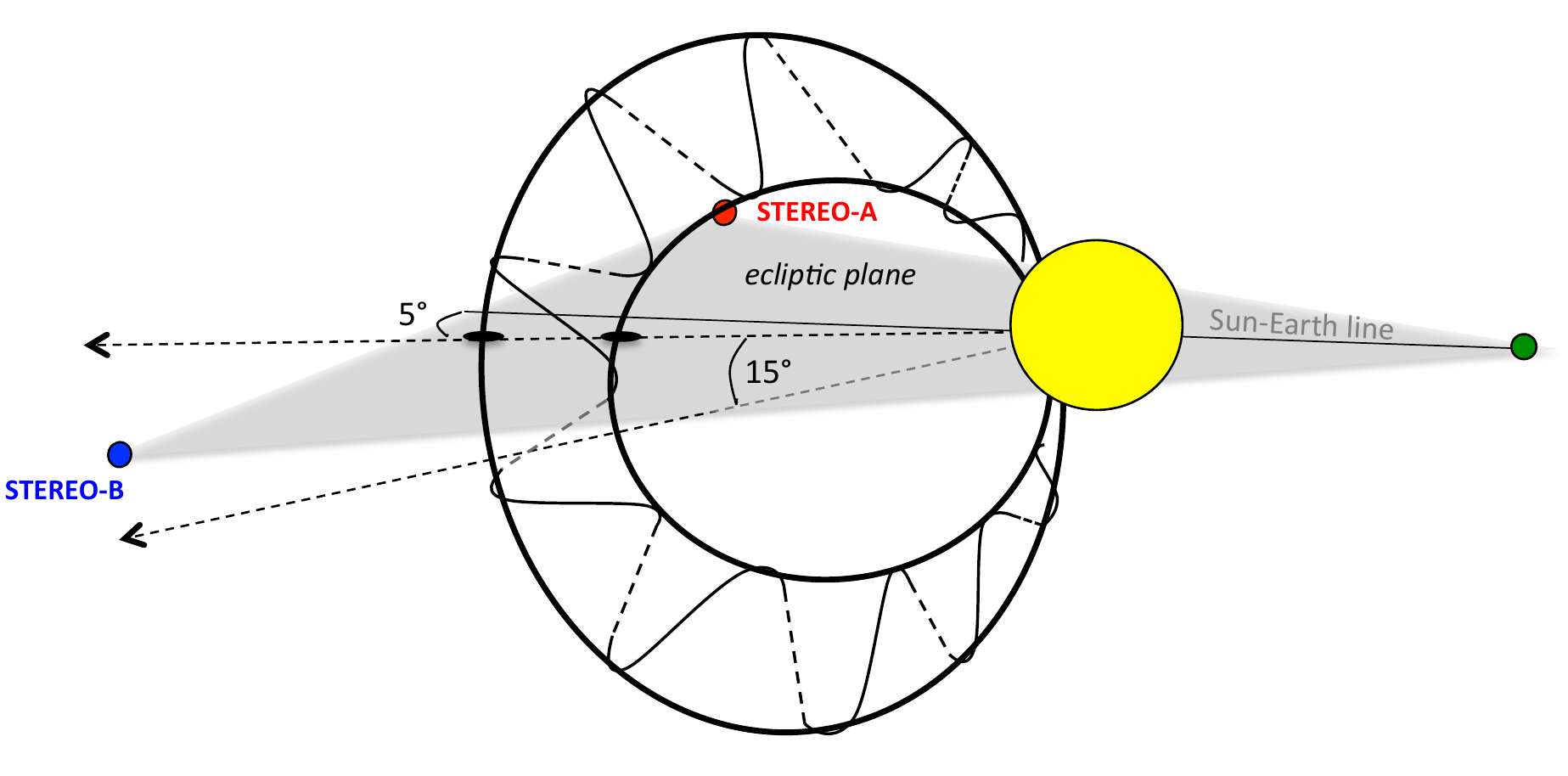}}
\end{minipage}%
\\
\begin{minipage}{\linewidth}
\centering
\subfloat[]{\includegraphics[width=0.85\textwidth, clip=true, trim = 0mm 50mm 0mm 0mm]{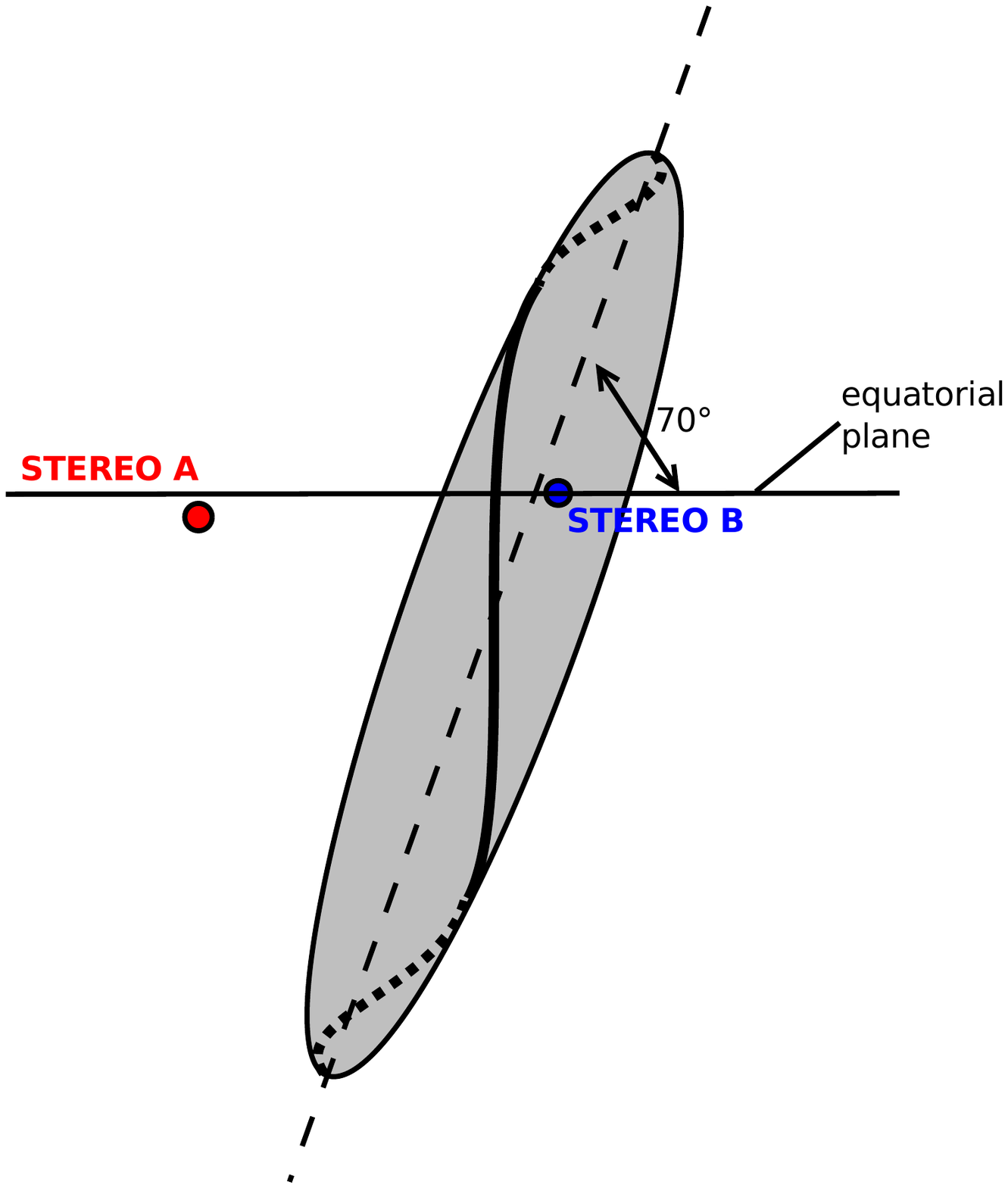}}
  \end{minipage}
  \caption{Cartoons showing the orientation and topology of the MC as determined by the GCS model with respect to STA (red dot) and STB (blue dot). (a) The propagation direction of the MC is towards the backside of the Sun as seen from Earth, 5$\degree$ away from the Sun-Earth line (E175) and 15$\degree$ to the south (as estimated by the GCS model close to the Sun). (b) View from outside 1\,AU back towards STA and STB. The MC front (gray ellipse) is strongly inclined with respect to the equatorial plane and nearly centrally passes STB but not STA (as estimated by the in situ model).} \label{fig:sketch_low_twist}
\end{figure}%
%
%
%
\subsection{Electron path lengths}\label{sec:path_length}
Solar energetic particles inside closed magnetic structures lend themselves as probes of the magnetic topology they propagate through.
If the time of the SEP injection at the Sun is known as well as the travel time and the energy of those SEPs,  their propagation path length  can be determined.
In terms of MCs this path length implies information on the twist of the magnetic field lines inside the structure.
A longer estimated path length therefore denotes a stronger twist (or longer loop legs).
As discussed in the previous section the SEP observations at STB strongly suggest that energetic electrons were injected separately into both loop legs of a magnetic cloud.
The observations in the NORTH and SOUTH SEPT telescopes of STB allow {two} different electron path lengths (one per loop leg) to be determined for the same MC at the same time.
We make the following assumptions:
\begin{itemize}
\item The two distinct radio sources seen in Fig. \ref{fig:radio2sources} represent the injections into the two MC loop legs. \\
This provides two injection times: the northern injection at 10:15:03\,UT and the southern injection at 10:16:33\,UT (the light travel time of 8.0 minutes to STA and 8.9 minutes to STB has been taken into account).
\item The near-relativistic electrons are injected simultaneously with the type III generating electrons.
\item The northern (southern) NRH source/injection is associated with electrons detected in the NORTH (SOUTH) SEPT telescope at STB.
\item The first arriving particles used to determine the onset times were propagating scatter-free.
\end{itemize}
The resulting path lengths of the analyzed 145-165\,keV electrons at STB are 2.38$\pm$0.08\,AU along the northern loop leg and 3.19$\pm$0.08\,AU along the southern loop leg.
The uncertainties of the path lengths are defined by the resolution of the SEPT data (1 minute) used to determine the onset times.
Furthermore, we consider the northern path length at STB as an upper limit because only unclear velocity dispersion is observed in the onset times in the NORTH telescope.
We suggest that STB enters a filled flux tube \citep[cf.][]{Borovsky2008} which might be associated with small changes in the magnetic field direction which coincide with the two peaks during the onset in the NORTH telescope at STB (Fig. \ref{fig:SEPT_sectored_intro}).
We note that the SOUTH telescope shows a clear velocity dispersion.
\\
Following the same approach as for STB and assuming a solar injection time at 10:15\,UT (at the time of the first type III burst) we determine a path length of 145-165\,keV electrons reaching STA of 2.24$\pm$0.08\,AU.\\
However, because  no SEPT telescope at STA covers the direction of the magnetic field during the onset of the event, the real onset might have been earlier.
Applying the SEP transport code {\it SEPinversion} \citep{Agueda2008, Agueda2012} provided by the SEPserver\footnote{\url{http://www.sepserver.eu}} database the onset time we determine is one minute earlier (at 10:35\,UT) at pitch angle 0$\degree$ .
The model furthermore reveals a radial mean free path 0.27\,AU and a later injection time at 10:21\,UT, which still lies within the duration of the series of type III bursts observed from 10:15 until $\sim$10:45\,UT.
Using the values found by the model of an SEP onset at 10:35\,UT and an injection time at 10:21\,UT, the path length reduces to 1.70$\pm$0.08\,AU. 
We therefore suggest that the uncertainty of the path length at STA is large with values lying between 1.70 and 2.24\,AU.
\\
A discussion of the electron path lengths in terms of the magnetic field topology in the MC (twist of magnetic field lines) follows in Section \ref{disc2:twist}.
%
\section{Magnetic cloud launched on 4 November 2013}\label{sec:mag_cloud}
\subsection{MC observations and modeling}\label{mc_obs}
The MC-like structure which embeded STB during the SEP event on 7 November 2013 arrived at the spacecraft on 6 November at 13:38\,UT and ended at 7 November 14:00\,UT (cf. ICME catalog \url{http://www-ssc.igpp.ucla.edu/forms/stereo/stereo_level_3.html}).
An inspection of the STEREO/COR1 and SOHO/LASCO coronagraph observations together with the CACTUS and LASCO CME catalogs reveals that this ICME is most likely associated with a CME launched at 4 November 2013 around 5\,UT.
The estimated speeds from the list and our own estimation (see below) fit the observed arrival time at STB well within a few hours. 
Furthermore, no other appropriate candidate CME is observed at that time.
The CME is associated with a flare that occurred on 4 November in the same AR as the event on 7 November.
From STB EUVI observations we derive the starting time of the associated flare which is around 4:15 UT. 
In order to derive the 3D morphology and average speed of the CME close to the Sun, we use the Graduated Cylindrical Shell (GCS) model by \cite{Thernisien2006, Thernisien2009}. 
The model is applied to white-light data from the COR2 instruments on board STA and STB as well as LASCO on board SOHO. 
The GCS uses a forward modeling technique in order to find the best match between the idealized model flux rope and the observed CME. 
The 3D kinematics reveals a speed of about 1250$\pm$50\,km/s at a distance of around 15\,R$_S$ from the Sun. 
The outcome of the model result regarding the orientation and topology is shown in Fig. \ref{fig:gcs_model}. 
The CME propagates in the direction E175S15, i.e., almost centrally between the two STEREO spacecraft, towards the south. 
We note that in comparison to the position of the associated flare at W168N3, the CME shows a slightly non-radial propagation towards STB and the south.
The CME face-on and edge-on width are found to be 100$\degree$ and 60$\degree$, respectively, and the inclination is found to be  -85$\degree$. 
This means that the CME is nearly completely oriented north-south . Typical uncertainties of these values are around 10$\degree$ (see, e.g., \cite{Mierla2010}). 
An estimation of the total loop length at 1\,AU reveals 3.53$\pm0.24$\,AU with a length of the northern (southern) leg of 1.33$\pm0.09$\,AU (2.20$\pm0.15$\,AU).
We note that a loop leg in this picture denotes the part of the magnetic cloud connecting STB with the Sun either along the northern leg of the MC or the southern leg, i.e., the MC is separated into two parts at the position of the equatorial plane.
The longer southern leg results from the southward tilt of the ICME.
\\
Figure \ref{fig:mc_model} shows in situ magnetic field and plasma measurements by STB (black dots) together with fits from the Global Magnetic Cloud (GMC) model \citep[][and references therein]{Hidalgo2014}.
The purpose of this model is to reconstruct the local orientation of the MC axis which is of certain interest because it is not known whether the ICME keeps its orientation (as determined by the GCS model close to the Sun) during its outward propagation.
The result of the fits shown in Fig. \ref{fig:mc_model} is a somewhat smaller inclination of 72$\degree$ of the structure compared to the GCS results.
Furthermore, it yields an impact parameter (distance to the MC axis) of 0.05\,AU and the estimated diameter of the MC at 1\,AU is 0.32\,AU.
The uncertainty of the model results is about 10\%.
The values suggest that the MC propagated almost directly towards STB instead of centrally between the two STEREO spacecraft as suggested by the GCS model close to the Sun.
The estimated diameter strongly suggests that STA does not observe the MC in situ.
However, because STB does not observe an ideal MC structure, the model may suffer larger uncertainties and strongly depends on the chosen temporal boundaries for the fitting.
\section{Results II: MC topology and orientation in the IP medium}\label{sec:results_mc}
Figure \ref{fig:sketch_low_twist} (a) shows a sketch of the MC orientation as described in the previous section together with the spacecraft positions.
A comparison of the GCS model results and the in situ MC modeling suggests that the MC exhibits less inclination at 1\,AU than close to the Sun.
Furthermore, the GCS model suggested a propagation of the ICME nearly centrally between the two STEREO spacecraft.
Although the in situ model shows that STB passes the structure fairly close to its axis, this does not necessarily denote a further non-radial propagation of the MC towards STB.
Because of the southward propagation of the ICME, together with its strong inclination, the nose of the structure may still be longitudinally situated centrally between the two STEREO spacecraft.
Fig. \ref{fig:sketch_low_twist} (b) shows a view from outside 1\,AU back towards both STEREO spacecraft.
The gray ellipse represents the strongly inclined front edge of the MC passing STB nearly centrally in terms of its axis, but not in terms of its nose.
The longitudinal separation angle between the STEREO spacecraft is 68$\degree$, which corresponds to a distance of $\sim$1\,AU between the two spacecraft. 
As discussed in Section \ref{mc_obs} the diameter of the MC as determined by the in situ modeling is 0.32\,AU.
Therefore, it is very unlikely that STA observes the same MC in situ.
\section{Discussion: Electron path lengths as a probe for the field line twist of the MC}\label{disc2:twist}
In addition to the models, SEP observations can provide an important tool for probing the topology of a magnetic cloud.
If the SEP injection time at the Sun is known, as well as the onset time at the observer, then one can determine the SEP propagation path length from the time delay and the known SEP energy as is done in Section \ref{sec:path_length}.
Because helical magnetic flux rope models such as the Lundquist model \citep{Lundquist1950} require a larger twist or winding of the magnetic field lines with increasing distance from the flux rope axis \citep[e.g.,][]{Lepping1990}, SEPs propagating through regions close to the flux rope borders are expected to have a significantly longer propagation path. 
\cite{Larson1997} determined electron path lengths of $\sim$4\,AU at the exterior of a MC.
\cite{Kahler2011a, Kahler2011b} statistically investigated ACE and Wind observations of $<100$\,keV electron events observed inside magnetic clouds and compared the estimated electron path lengths to model predictions.
Interestingly, they rarely found these significantly longer path lengths close to magnetic cloud borders and often no difference between path lengths inside and outside MCs were found.
\cite{Hu2015} re-investigated the set of events observed by Wind provided by \cite{Kahler2011b} and found a reasonable correlation between the electron path lengths and results of the Gold-Hoyle model of a constant twist (see \cite{Hu2015} and references therein) suggesting that the twist of the magnetic field lines inside MCs may be much smaller than predicted by the Lundquist model.\\
The SEP event on 7 November 2013 occurred less than four hours before STB exited the MC-like structure.
As discussed in Section \ref{sec:results_mc} the MC-like structure passed STB nearly centrally; however, the 7 November SEP event occurred $\sim$4 hours before the spacecraft exited the MC,  suggesting that STB was situated close to the MC border at that time.
\\
The special situation of two separately injected beams into the MC that reach STB along the different loop legs provides the opportunity to determine two distinct electron path lengths for the same SEP event.
In Section \ref{mc_obs} the results of the GCS model were used to estimate the length of the whole loop (3.53$\pm0.24$\,AU), as well as the lengths of the northern (1.33$\pm0.09$\,AU) and southern (2.20$\pm0.15$\,AU) loop legs connecting to STB.
\\
As shown in Section \ref{sec:path_length} we find an electron path length of 2.38$\pm$0.08\,AU (3.19$\pm$0.08\,AU) along the northern (southern) loop leg for STB yielding a total path length of 5.57\,AU.
The longer path length along the southern loop leg is in agreement with the southward orientation of the ICME causing unequal loop leg lengths for an observer in the equatorial plane. 
We note that in the case of the northern loop leg we consider the calculated path length as an upper limit because the onset at the spacecraft occurs simultaneously over all energies suggesting that STB enters an already filled flux tube (see Section \ref{sec:path_length}).
An earlier onset would have caused a smaller path length.
The estimated electron path lengths are hence around 50\% longer than the loop legs length inferred from the GCS model suggesting a moderate amount of field line winding at the exterior of the MC.
Furthermore, the in situ magnetic field direction observed by STB during the passage of the MC is nearly perfectly aligned with the N-axis (in RTN coordinates) pointing towards the north.
Together with the high inclination of the MC this shows that the in situ magnetic field instead points along the magnetic cloud axis, again suggesting a weak magnetic field winding.
We therefore find our results in agreement with \cite{Kahler2011a, Kahler2011b} and \cite{Hu2015}, who conclude that the field line twist must be fairly small.
Following \cite{Kahler2011a} for a rough estimation of the number of field line rotations N and the magnetic field twist $\tau$ in units of radians per AU inside the MC we determine N$\approx$4 and $\tau\approx 8,$ which lies inside the range found by \cite{Kahler2011a} for the whole MC loop.
We note that if we determine these values separately for the two loop legs, we obtain N$\approx$2 and  $\tau\approx 9$ for the northern loop leg and N$\approx$2 and  $\tau\approx 7$ for the southern loop leg, respectively.\\
For comparison we also determine the electron path length for STA, which is most likely situated in an open magnetic field structure during the onset of the event.
Assuming an injection time at 10:15\,UT when the first type III radio burst is observed (and an SEP onset time at 10:36\,UT) we receive a path length of 2.24\,AU.
The reason for this rather long path length could be a strongly distorted magnetic field due to the preceding SIR (see Section \ref{sec:in_situ}).
However, a series of type III bursts is observed with a second one at 10:21\,UT. 
Assuming this one as the associated injection and a one minute earlier onset time at 10:36\,UT as derived from the transport model (see {Section \ref{sec:path_length}) results in a smaller path length of 1.70$\pm0.08$\,AU.\\
%
%
%
\section{Conclusions}
The SEP event on 7 November 2013 was observed by both STA and STB, which were longitudinally separated by 68$\degree$ at that time. 
While STA is nominally well connected to the source AR and observes an anisotropic SEP beam streaming from the Sun as expected, STB observes a bi-directional distribution of near-relativistic electrons in the north-south direction.
In situ plasma and magnetic field observations show that STB is embedded in a magnetic cloud like structure during the onset and first $\sim$4 hours of the event.
We were able to identify that the parent CME was launched on 4 November 2013 from the same AR that produced the SEP event on 7 November.
A combination of observations and modeling yields a non-radial propagation of the CME towards the south and STB so that if finally passes STB nearly centrally at 1\,AU.
The MC exhibits a high inclination of 85$\degree$ close to the Sun (results of the GCS model) and keeps a high inclination of $\sim$70$\degree$ while propagating to 1\,AU (results of the GMC Model).
Fig. \ref{fig:sketch_low_twist} shows cartoons illustrating the north-south oriented MC and spacecraft constellations.
\\
The main results of the present study are:\\
\begin{itemize}
\item{\bf The bi-directional SEP distribution observed by STB is produced by two distinct injections into both loop legs of the MC, which is still anchored at the Sun and embeds STB.}
Although the first arriving particles are observed in the NORTH telescope of SEPT, the beam that arrives 12 minutes later  in the SOUTH telescope is more intense than the first beam (see Fig. \ref{fig:SEPT_sectored_intro}; we note that the factor is energy-dependent and at $>=$200\,keV the intensities in the NORTH and SOUTH telescopes are comparable.)
The relative intensity and timing cannot be explained by a mirroring scenario where the mirrored beam usually exhibits less than 30\% of the incident intensity \citep{Agueda2010, Wang2011, Klassen2012}. 
Furthermore the delay between the two onsets is much shorter than expected from mirroring {close to the Sun}.
However, a mirroring point at a distance of $\sim$0.5\,AU to STB would fit the delay.
Radio observations by the Nancay Radioheliograph (see Fig. \ref{fig:radio2sources})  further support the above scenario by showing two distinct radio sources at the same time, which are likely associated with the two separate injections into the MC loop legs.\\
\item {\bf The injection into both loop legs requires an extended injection region that is most likely provided by the associated coronal shock,} which is indicated by the associated type II radio burst (see Fig. \ref{fig:radio}). 
However, this extent depends on how far the loop legs of the CME are separated at the injection height;   diverging magnetic field lines low in the corona could also provide this extent.\\
\item {\bf The winding of the magnetic field lines at the borders of the magnetic flux rope {is found to be of moderate size.}}
The calculated electron path lengths corresponding to the event observed in the NORTH (SOUTH) SEPT telescope at STB of 2.38\,AU (3.19\,AU) are around 50\% longer than the modeled lengths of the loop legs (see Section \ref{disc2:twist}).
These results are in agreement with \cite{Kahler2011a, Kahler2011b} and \cite{Hu2015}, suggesting that the twist of magnetic field lines at the exterior of MCs is smaller than predicted by the Lundquist model.
Furthermore, the in situ magnetic field direction measured by STB throughout the MC encounter points nearly perfectly towards the north, i.e., in the same direction as the orientation of the MC itself.
This is also in agreement with only small magnetic field line winding at the border of the MC.
\\
The estimated length of the southern loop leg (based on the GCS model) and the electron path length along the southern loop are significantly longer than for the northern loop leg.
This is in agreement with the southward orientation of the MC causing the larger part of the MC to be situated in the southern hemisphere. 
\end{itemize}
This study highlights the importance of directional SEP observations. 
While a high resolution in pitch angle space as provided by STEREO/LET in the ecliptic plane is desirable in order to determine the width of a beamed distribution or a loss cone, viewing directions out of the ecliptic plane are indispensable for  analyzing observations like those of the 7 November 2013 event.
Directional SEP observations out of the ecliptic plane are exclusively provided by the SEPT instrument on board STEREO.
%
%
\begin{acknowledgements}
We acknowledge the STEREO PLASTIC, IMPACT, and SECCHI teams for providing the data used in this paper. 
The STEREO/SEPT Chandra/EPHIN and SOHO/EPHIN project is supported under grant 50OC1302 by the Federal Ministry of Economics and Technology on the basis of a decision by the German Bundestag.
This survey is generated and maintained at the Observatoire de Paris by the LESIA UMR CNRS 8109 in cooperation with the Artemis team, Universities of Athens and Ioanina, and the Naval Research Laboratory.
M.T. and A.V. acknowledge the Austrian Science Fund (FWF): V195-N16 and P24092-N16. 
\end{acknowledgements}

\bibliographystyle{aa.bst}
\bibliography{references}  

\end{document}